\documentclass[ reprint,  amsmath,amssymb,  aps, ]{revtex4-1}
\usepackage[dvipdfmx]{graphicx}
\usepackage{dcolumn}
\usepackage{bm}
\usepackage{amsmath}
\usepackage{amsfonts}
\usepackage{amssymb}
\usepackage{color}
\usepackage{comment}
\providecommand{\U}[1]{\protect\rule{.1in}{.1in}}
\begin{document}

\preprint{APS/123-QED}

\title{Universality of density of states in configuration space}
\author{Tetsuya Taikei}
\author{Kazuhito Takeuchi}
\author{Koretaka Yuge}
\affiliation{Department of Materials Science and Engineering, Kyoto University, Kyoto 606-8501, Japan}

\begin{abstract} 
 In this study, we confirm the universality of density of microscopic states in non-interacting system; this means statistical interdependence is vanished in any lattices.
This enable one to obtain information of configuration of solute atoms, free energy, phase diagram with performing first-principles calculation on few special microscopic states combined with our established theory.
\end{abstract}
\maketitle

\section{\label{sec:level1}Introduction} 
 For materials design, first we need to know constituent elements with phase diagram, and composition of each elements.
For clarifying the properties of materials more in detail, we need to reveal the configuration of solute atoms, which means the importance of quantitative deal with configuration space.
Especially in areas of computational materials science, the so-called Generalized Ising Model (GIM) \cite{GIM} has been widely used to describe configuration space with first-principles calculations. 
In the GIM, the configuration properties are specified with a complete set of coordination (i.e., basis functions), \{${q_1,q_2,....q_g}$\}.
However, the conventional approaches [][] to get information of configuration space of equilibrium state with GIM and Monte Carlo (MC) \cite{MC1}\cite{MC2} simulation needs a lot of first-principle calculations for every selected material even with the same lattice \cite{select1}\cite{select2}: they do not focus on how spatial constraint (i.e., lattice) plays an important role in equilibrium state.
\\
\ Recently, we have established new approach combined with GIM to address the problem of conventional approach:
With the new approach in composition-fixed system, we find that canonical average of each physical quantity $Q_u(T)$ is determined by configurational energy of 'Projection State' (PS) for each physical quantities \cite{EMRS2}. 
In composition-unfixed system, we obtain the fact that energy of PS for the coordinate describing composition ('Grand Projection State', GPS) determines connection between composition and chemical composition. Especially in binary system, GPS determines free energy, which enables one to obtain binary phase diagram for phase separation system with two GPSs \cite{EMRS4}.
Here, we emphasize that PS (including GPS) is only dependent on spatial constraint  (e.g., lattice for crystalline solids, volume and density for liquid in rigid box), independent of constituent elements, temperature, and interactions: One can know PS a priori only with the information of spatial constraint. 
This result, therefore, is not only (i) theoretically interesting in describing free energy by a single microscopic state, GPS, even though free energy has a member of entropy dependent on all possible microscope states,
but also (ii) practically very useful for the material design with avoiding large number of first-principles calculations; one need to perform first-principle calculations on few structures for estimating physical quantities/free energy, and constructing phase diagram. 
\\
These approach is based on the 'statistical independence' for $\{q_u\}$ on a non-interacting system in thermodynamical limit for the number of atoms even though the basis
functions themselves are not essentially statistically independent: Here, statistical independence means that (A) ideally numerical vanishment of non-diagonal elements of covariance matrix for $\{q_u\}$ and  
(B) the density of microscopic states for $\{q_u\}$ on a non-interacting system can be well characterized by a multi-dimensional Gaussian distribution. 
However, the validity of this assumption is confirmed only in FCC lattice in composition-fixed system (CFS) and in composition-unfixed system (CUFS) \cite{EMRS4}\cite{EMRS5}.
In this study, we confirm the universality of density of states in configuration space with random matrix (RM) constructed by Gaussian orthonormal ensemble through the MC simulation on real lattice, which means the validity of our assumption in any lattices.
In CFS, we obtain the connection of the statistical interdependence and the number of atoms, $N$ numerically.
In CUFS, we demonstrate that statistical interdependence remains however we make another lattice from one, which certainly indicate universality of density of states.
In the following, we demonstrate the validity first in CFS in order to confirm that PS can describe $Q_u(T)$ in any lattices, then in CUFS in order to confirm that GPS can describe free energy in any lattices.
%
\section{\label{sec:level2}Methodology and Discussions}
 Let us first explain how we have confirmed  the statistical interdependence in FCC lattice \cite{EMRS5}.
 For considering density of states on configuration space, we take the matrix $A$.
We construct $A$ with its $(s,t)$ element $a_{st}$ denotes the value of $q_t$ at sampling time $s$.
 Therefore, when we sample $m$ points on the configuration space and consider $n$ kinds of basis functions, $A$ should be $m \times n$ matrix.
With $A$, we construct the covariance matrix $R$;
\begin{equation}
 R=\frac{1}{m}A^{\rm{T}}A,
 \label{eq:matrix}
\end{equation}
where  $A^{\rm{T}}$ means transposed matrix of $A$. 
When we consider an ideal system where statistical interdependence is disappeared and the distribution of the elements at each column is constructed by a Gaussian distribution, $A$ should be random matrix, $A_{\rm{RM}}$, with a Gaussian orthonormal ensemble.
Therefore, our strategy is to compare matrix constructed from the practical system and random matrix. 
In this study, we construct $A_{\rm{RM}}$ with its all elements independently consist of normal random numbers, with the average and variance respectively taking 0, and 1.
For comparison, $A$ of practical system is normalized so that the average and variance of each column respectively should be 0, and 1.
When we compare matrices, we focus on the density of eigenstates (DOE) of $A$ in order to avoid excessive estimation of numerical error in the simulation to sample from large configuration space.
For more quantitative comparison of DOEs, we estimate moments (from 2nd to 4th) of DOEs, defined by 
$M_1=\Sigma_{i=1}^{N_d} x_i/N_d$,
$M_L=\left\{\Sigma_{i=1}^{N_d} (x_i-M_1)^L/N_d\right\}^{1/L}$, 
where $L$ is an order of the moment, $N_d$ is number of data and $x_i$ is each data of index $i$.
\\
\ For setting calculation condition of each lattice, such as what kind of basis functions we consider, we take care about how one lattice (named Son Lattice, SL) is made from other lattice (named Mother Lattice, ML).
We determine kinds of considered clusters of SL with considering how the basis functions of ML is convereted to that of SL;
this connection makes considered clusters of SL more than that of ML.
We change sampling times, $M_{\rm{SL}}=M_{\rm{ML}}\times\frac{N_{SL}}{N_{\rm{ML}}}$;
this change in sampling times is appropriate for random matrix too.
Marchenko and Pasture showed that the DOE of covariance matrix for random matrix can be analytically determined with $m\rightarrow\infty$ when $n/m$ is fixed \cite{Marchenko}.
In this study, we prepare two sets of ML and SL, which Table.~\ref{HSL_LSL} shows.
The detail of how we make SL from ML and what kind of clusters we consider in each set, and the difference between BCC1 and BCC2 are shown in Appendix A.
\begin{table}
  \begin{center}
         \begin{tabular}{|l|l|l|}\hline 
         &ML& SL \\ \hline
        set1&FCC&BCC1,Diamond \\ \hline
        set2&BCC2&HCP\\ \hline
            \end{tabular}
         \caption{Sets of HSL and LSL．BCC1 and BCC2 are different in terms of considered clusters.}
         \label{HSL_LSL}
  \end{center}
\end{table}
\\
\ In this study, we consider an example for equiatomic A-B binary system on each lattice.
We employ generalized Ising spin model with spin variables of $\sigma = {\pm}1$ in order to get $\{q_u\}$.
Here, $q_u$ can be defined as  $q_u=\langle\prod_{i{\in}u} \sigma_i\rangle_{\rm{lattice}}$, where $\sigma_i$ is spin at site $i$, $\langle\cdot\cdot\rangle_{\rm{lattice}}$ is average over all sites on the lattice, and $u$ is the index indicating cluster type, such as empty, point, 1st nearest neighbor pair.
With this definition, we can take advantage to get complete orthonormal basis functions; this means we can expect (A) (explained in introduction), numerical vanishment of non-diagonal elements, without transforming coordination of basis functions.
In order to demonstrate the validity of our PS/GPS approach in CFS/CUFS, we prepare two types of covariance matrices corresponding to the individual conditions, which we explain more in detail in the following.
\subsection{\label{sec:level21}Universality of density of states in CFS}

\begin{figure*}[p]
  \begin{center}
    \begin{tabular}{c}
%
      \begin{minipage}{0.33\hsize}
        \begin{center}
          \includegraphics[width=\linewidth]{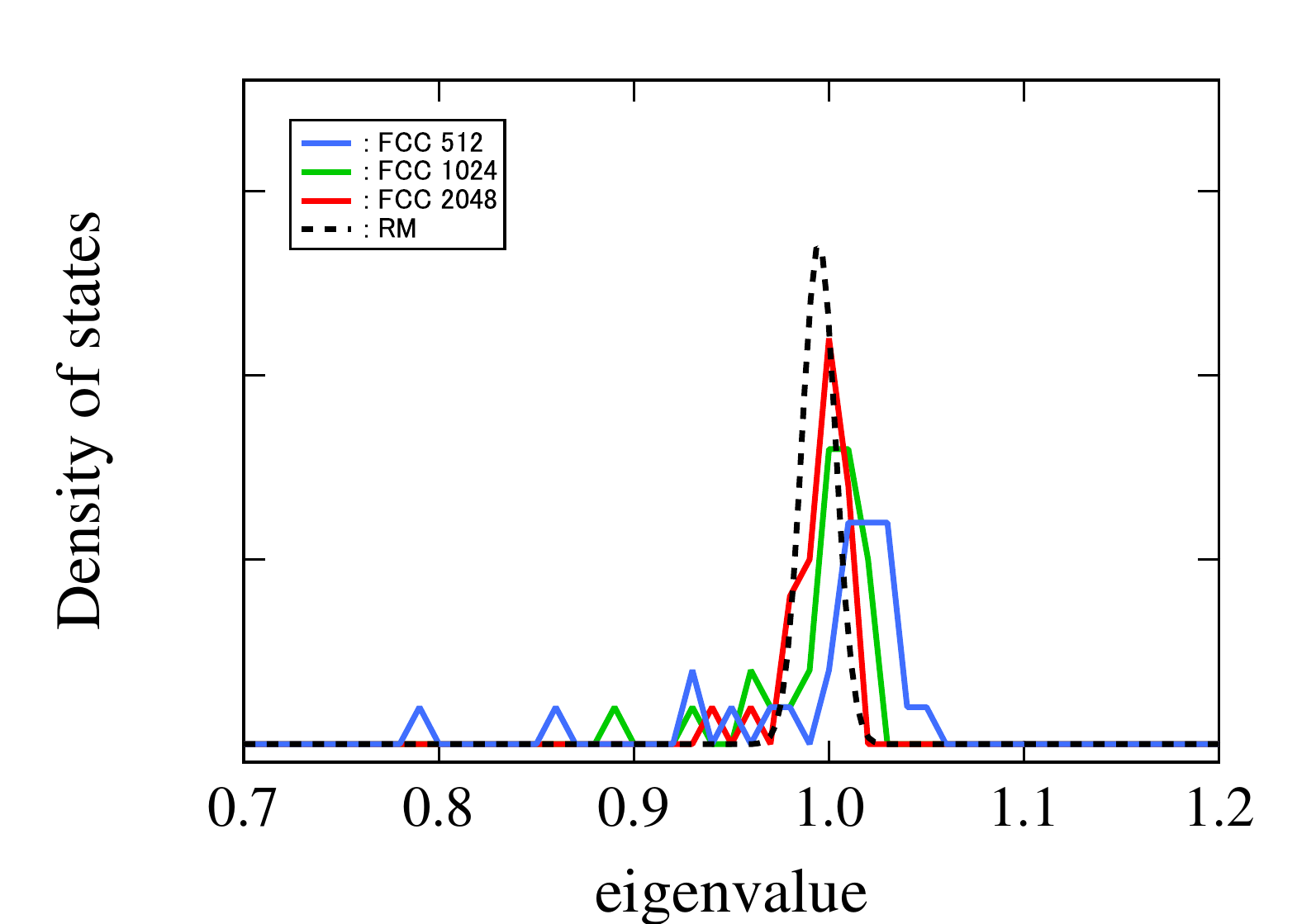}
        \end{center}
      \end{minipage}
      \begin{minipage}{0.33\hsize}
        \begin{center}
          \includegraphics[width=\linewidth]{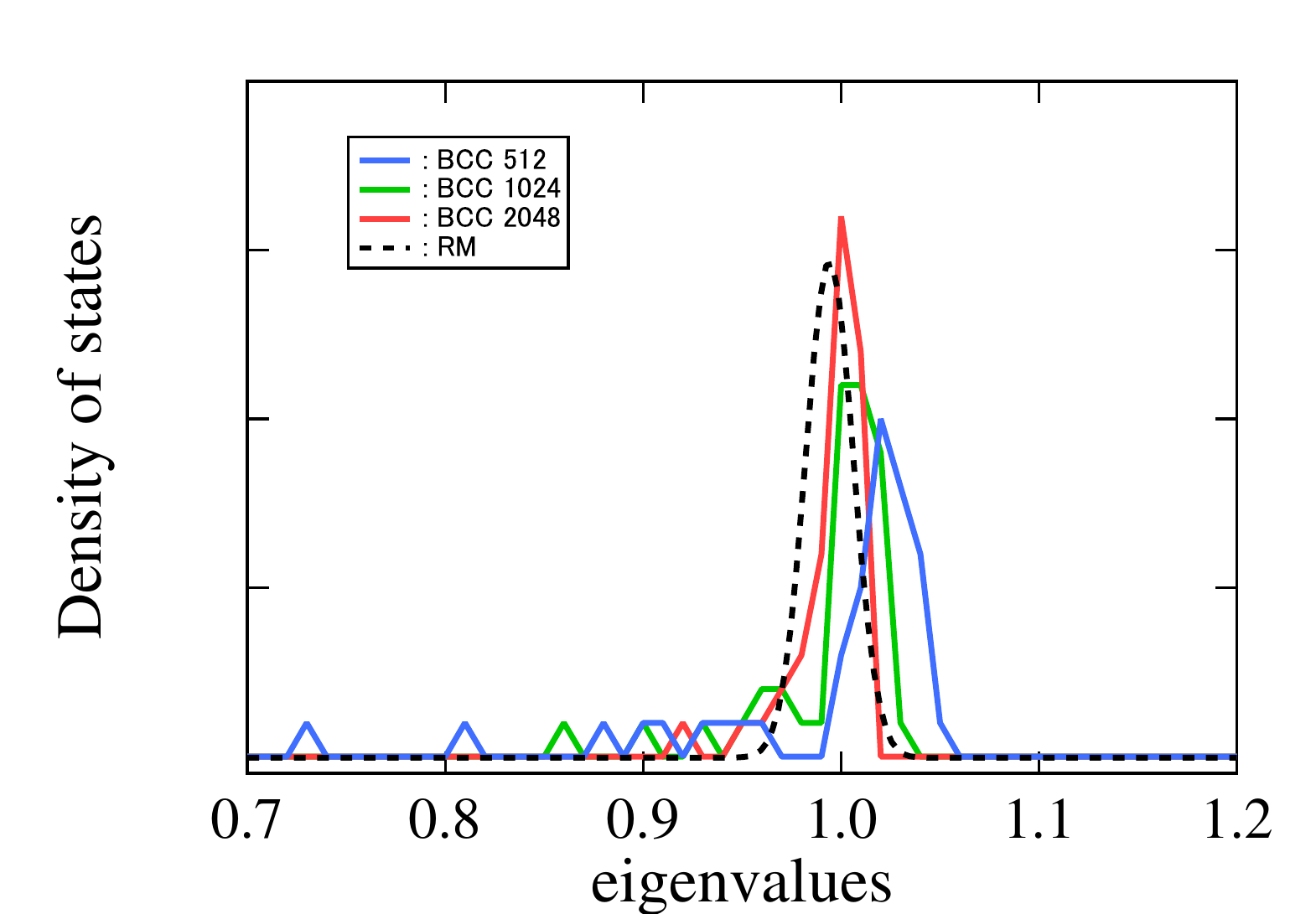}
        \end{center}
      \end{minipage}
\begin{minipage}{0.33\hsize}
        \begin{center}
          \includegraphics[width=\linewidth]{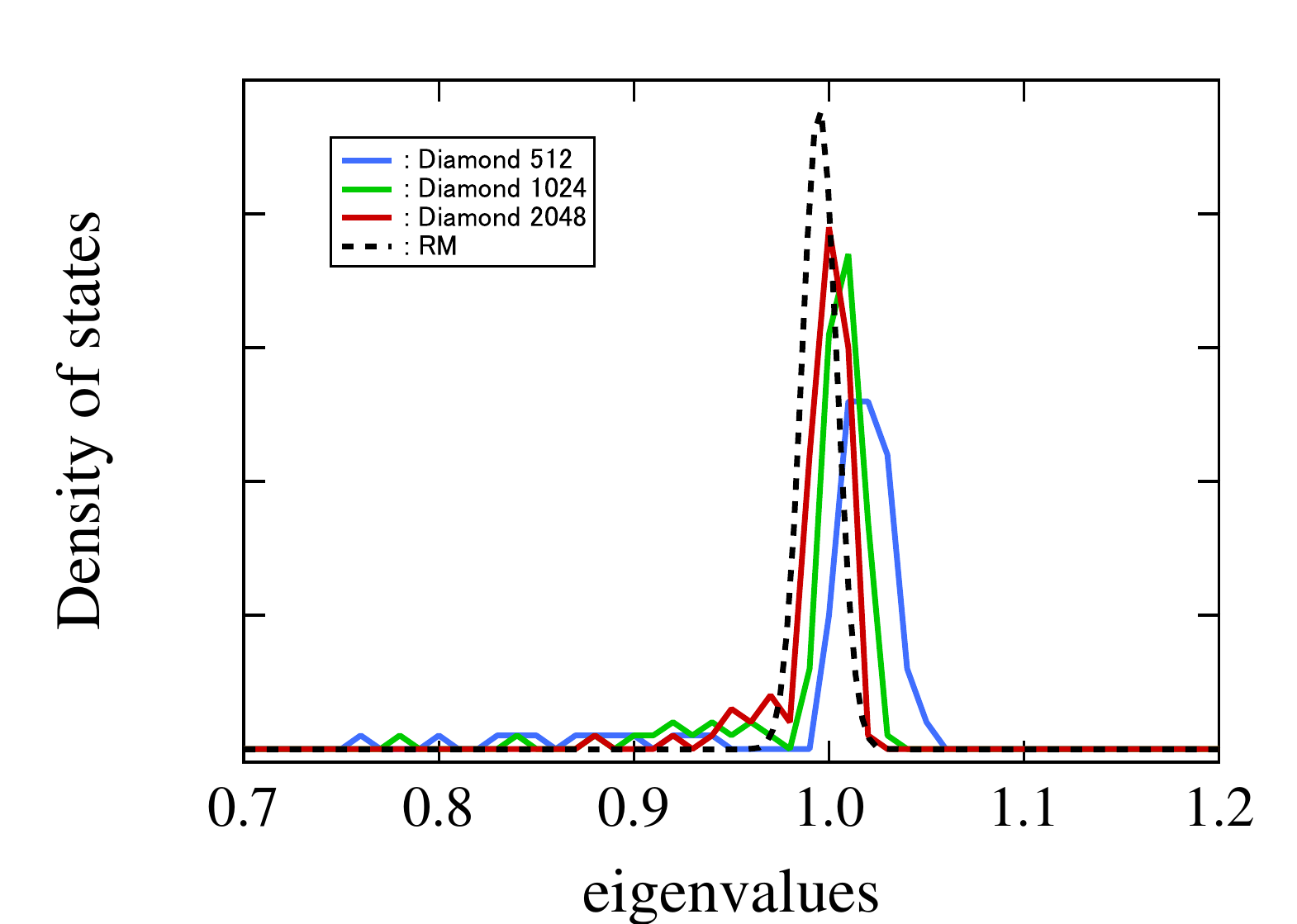}
        \end{center}
      \end{minipage}\\
\begin{minipage}{0.33\hsize}
        \begin{center}
          \includegraphics[width=\linewidth]{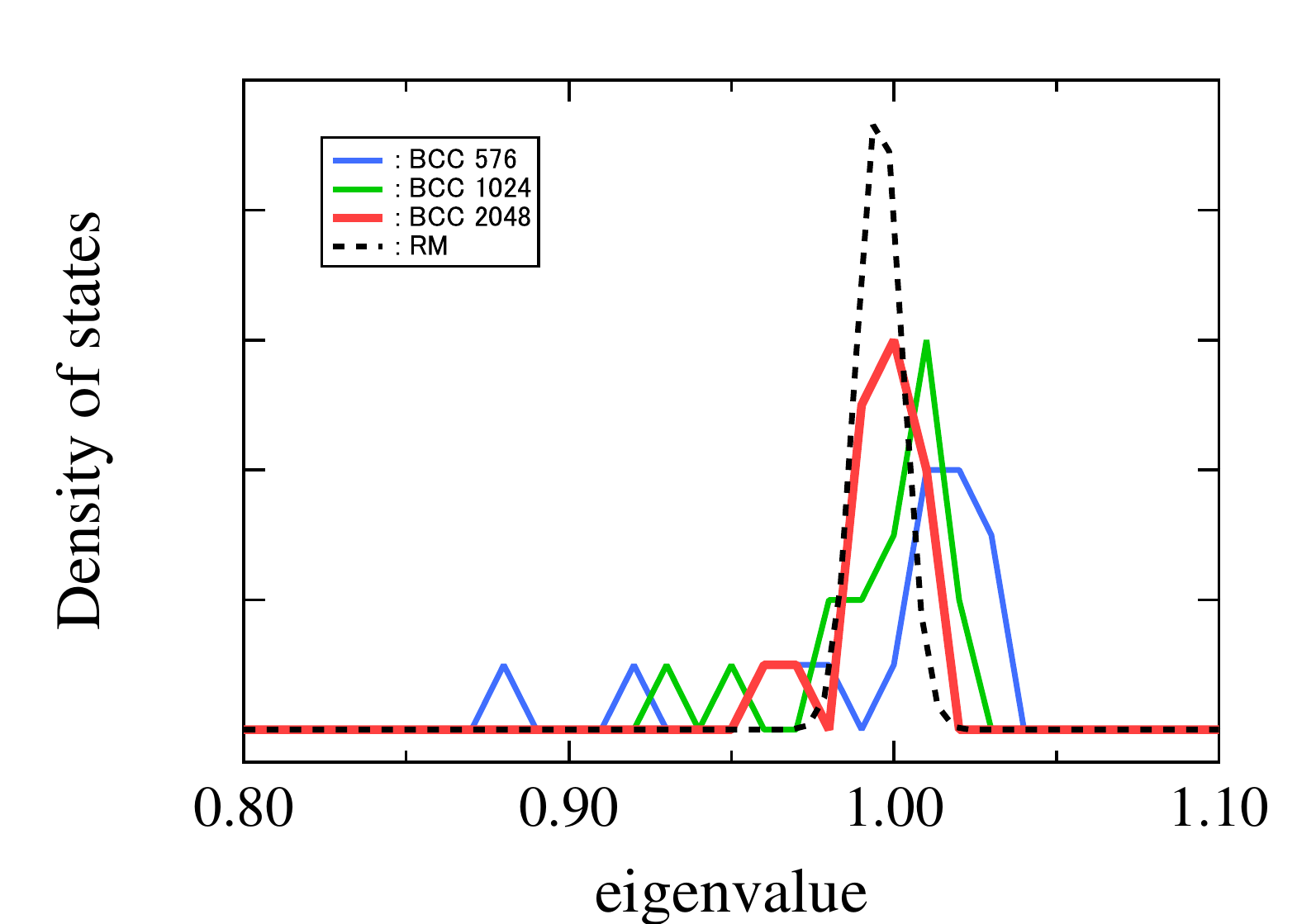}
        \end{center}
      \end{minipage}
\begin{minipage}{0.33\hsize}
        \begin{center}
          \includegraphics[width=\linewidth]{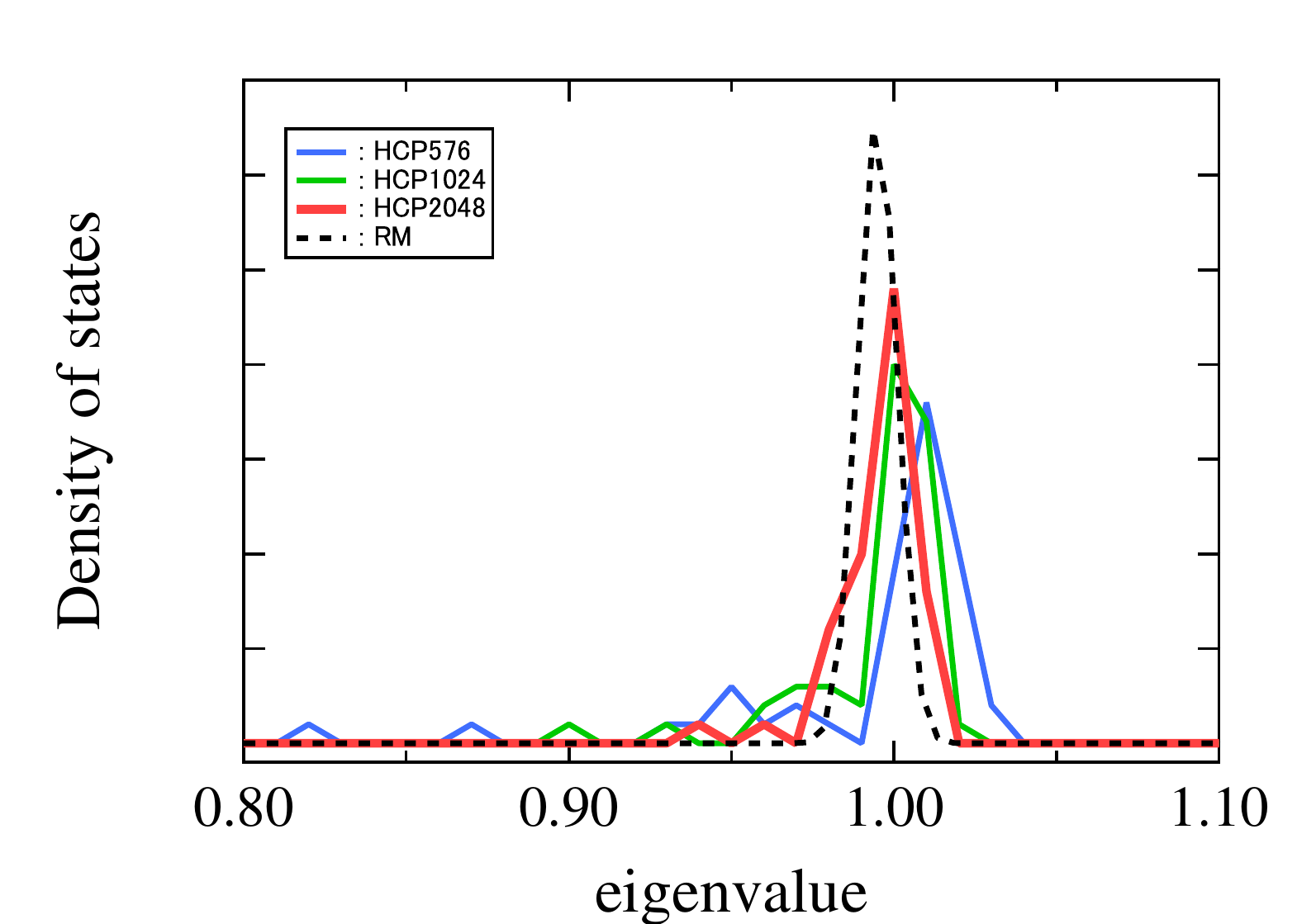}
        \end{center}
      \end{minipage}
    \end{tabular}
    \caption{ Up: DOEs for set1. Down: DOEs for set2. \\
 This landscape shows that statistical interdependence is confirmed with increase of the lattice size.}
    \label{fig.doe_all}
  \end{center}
\end{figure*}

\begin{figure*}
  \begin{center}
    \begin{tabular}{c}
%
      \begin{minipage}{0.33\hsize}
        \begin{center}
          \includegraphics[width=\linewidth]{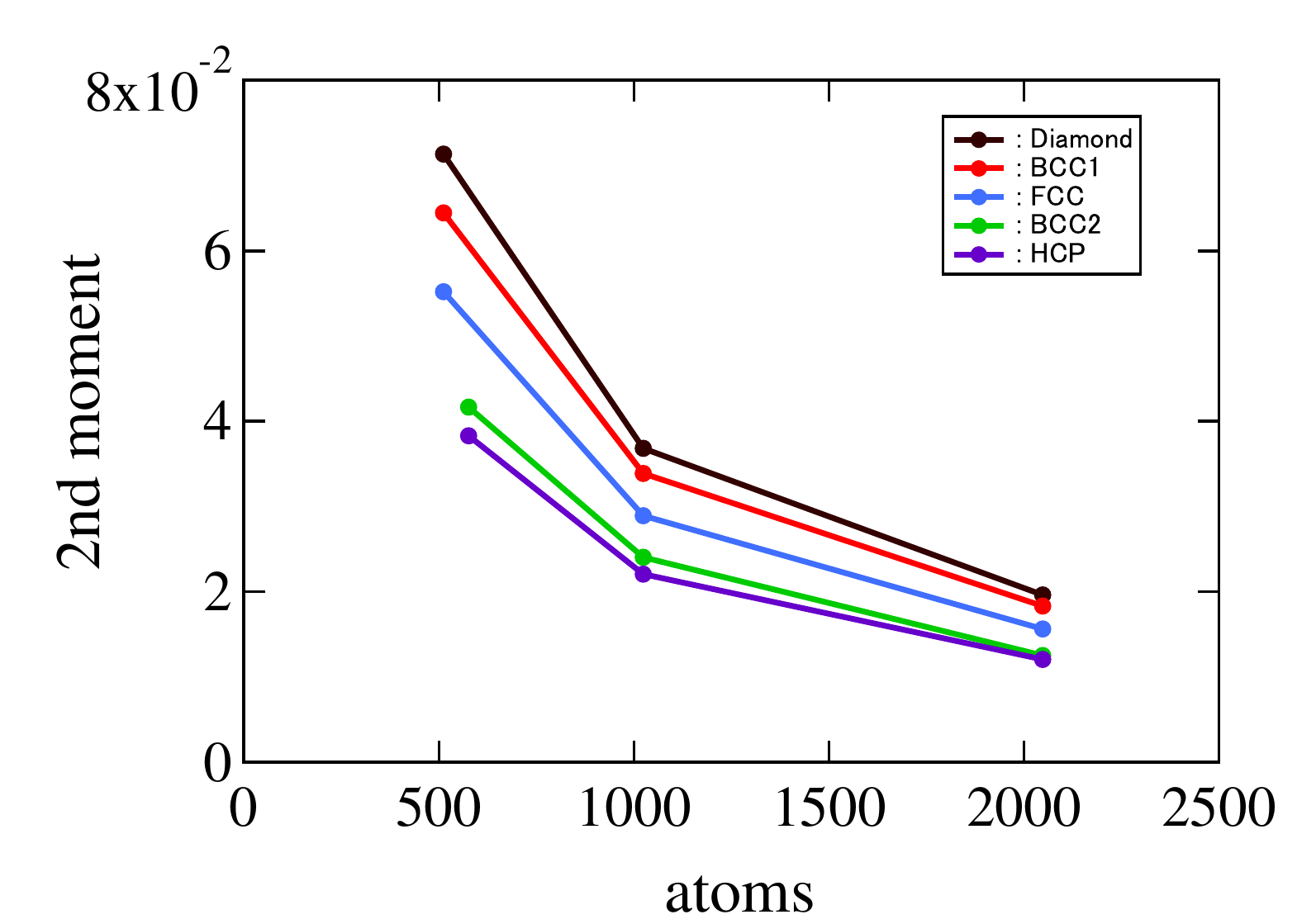}
        \end{center}
      \end{minipage}
      \begin{minipage}{0.33\hsize}
        \begin{center}
          \includegraphics[width=\linewidth]{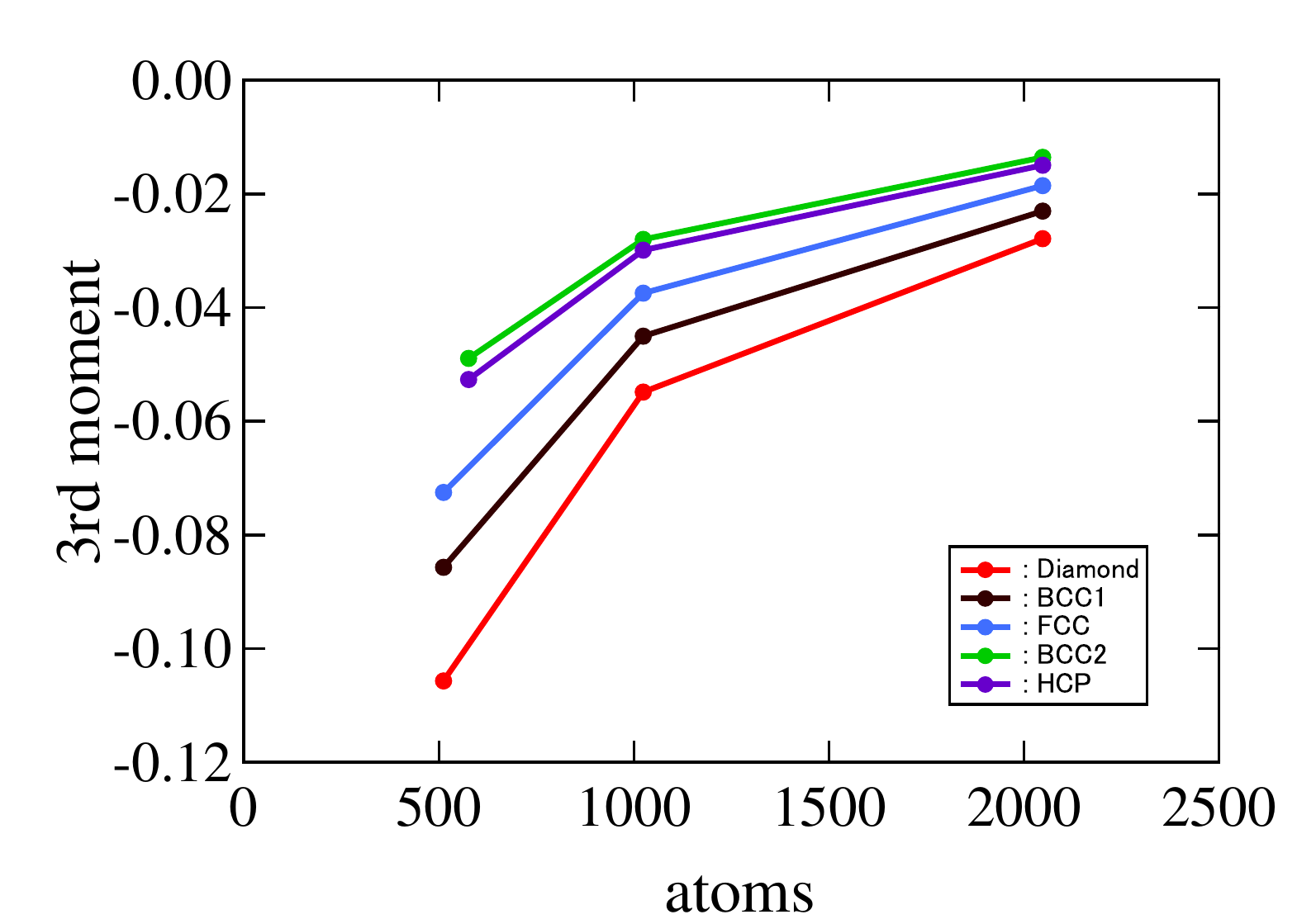}
        \end{center}
      \end{minipage}
\begin{minipage}{0.33\hsize}
        \begin{center}
          \includegraphics[width=\linewidth]{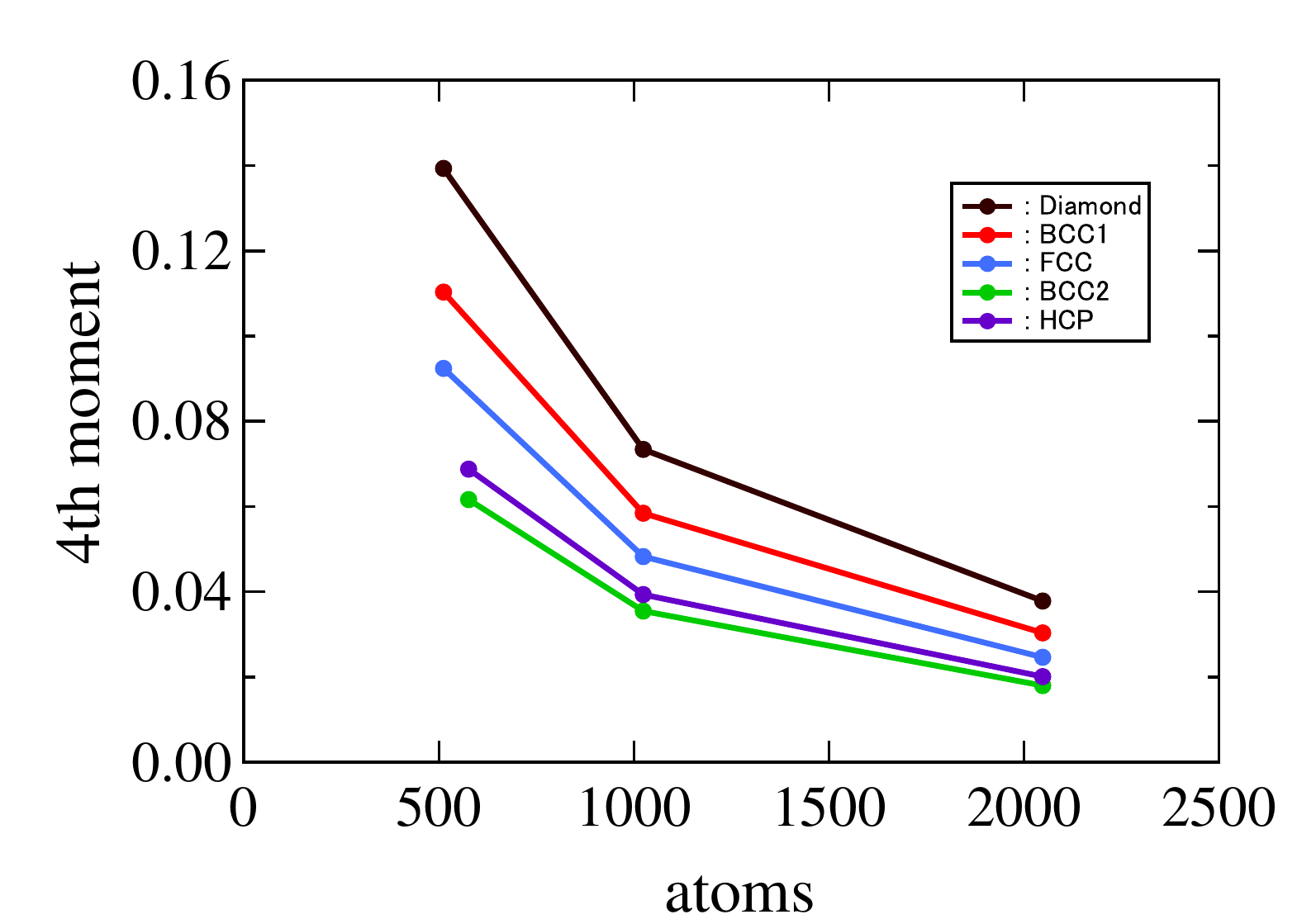}
        \end{center}
      \end{minipage}
    \end{tabular}
    \caption{2nd, 3rd, and 4th order moments of DOEs for CFS along the number of atoms. These shows that with increase of the number of atoms, statistical independence is guaranteed.}
    \label{fig.moment_all}
  \end{center}
\end{figure*}

\begin{figure*}
  \begin{center}
    \begin{tabular}{c}
%
      \begin{minipage}{0.33\hsize}
        \begin{center}
          \includegraphics[width=\linewidth]{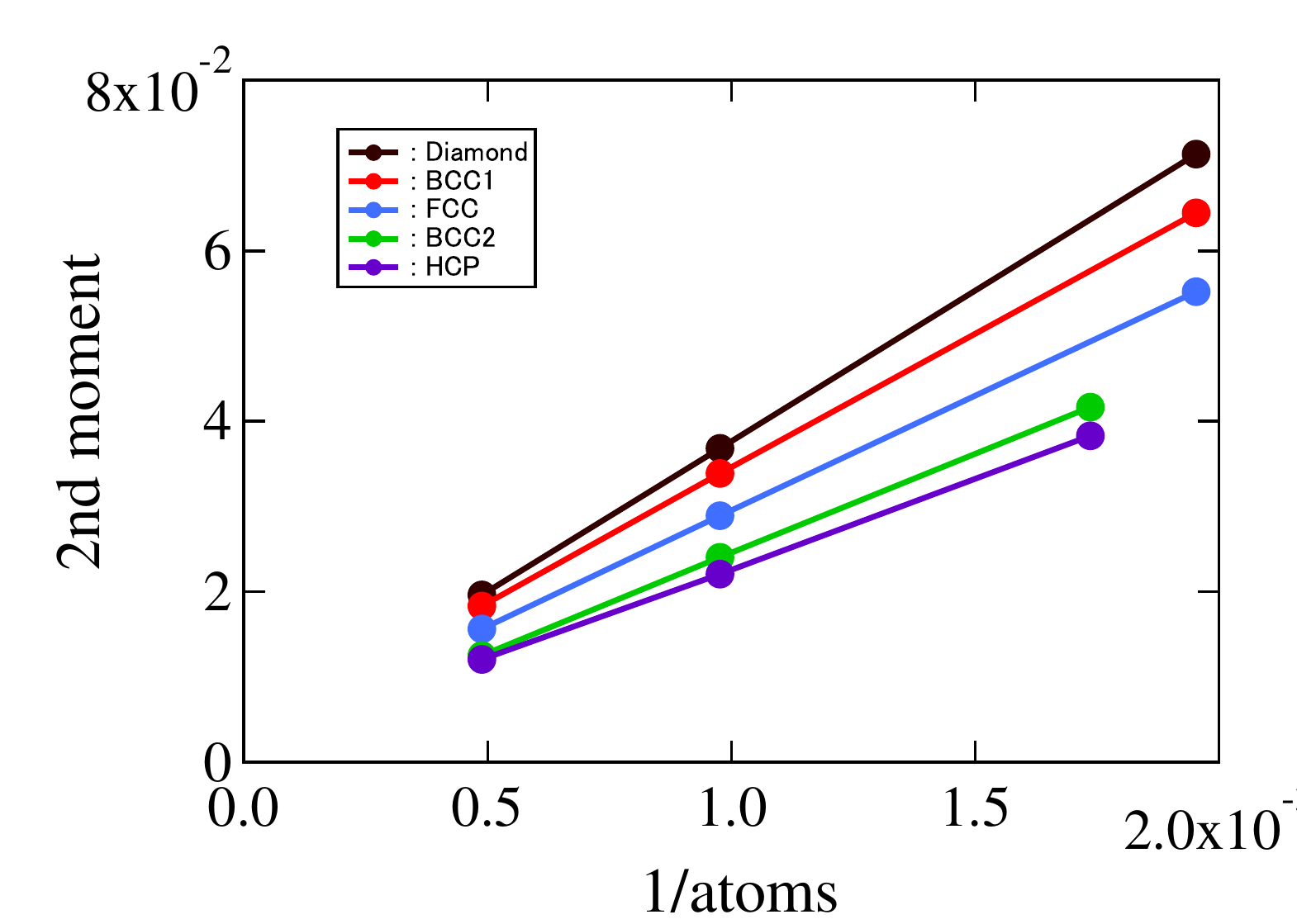}
        \end{center}
      \end{minipage}
      \begin{minipage}{0.33\hsize}
        \begin{center}
          \includegraphics[width=\linewidth]{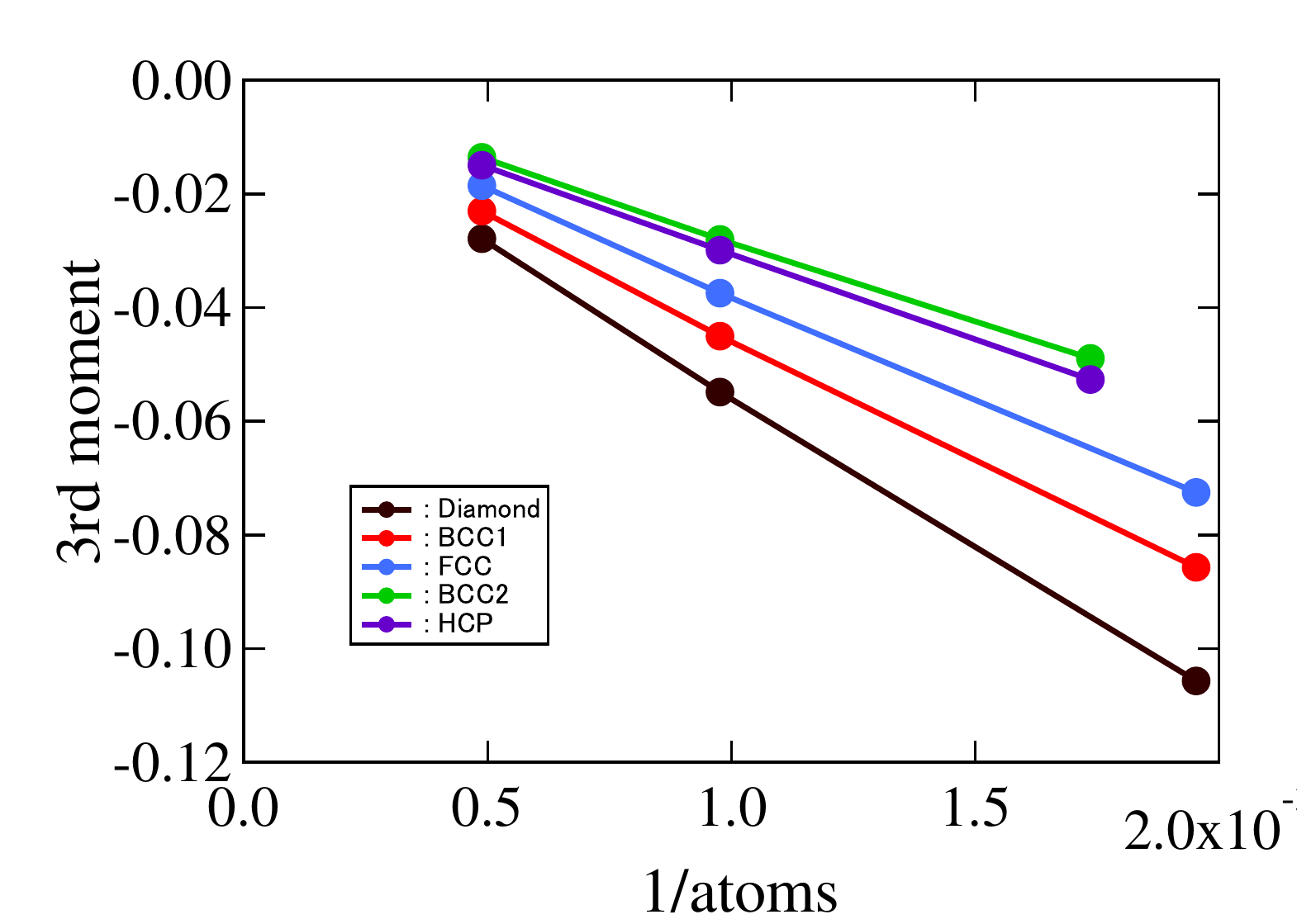}
        \end{center}
      \end{minipage}
\begin{minipage}{0.33\hsize}
        \begin{center}
          \includegraphics[width=\linewidth]{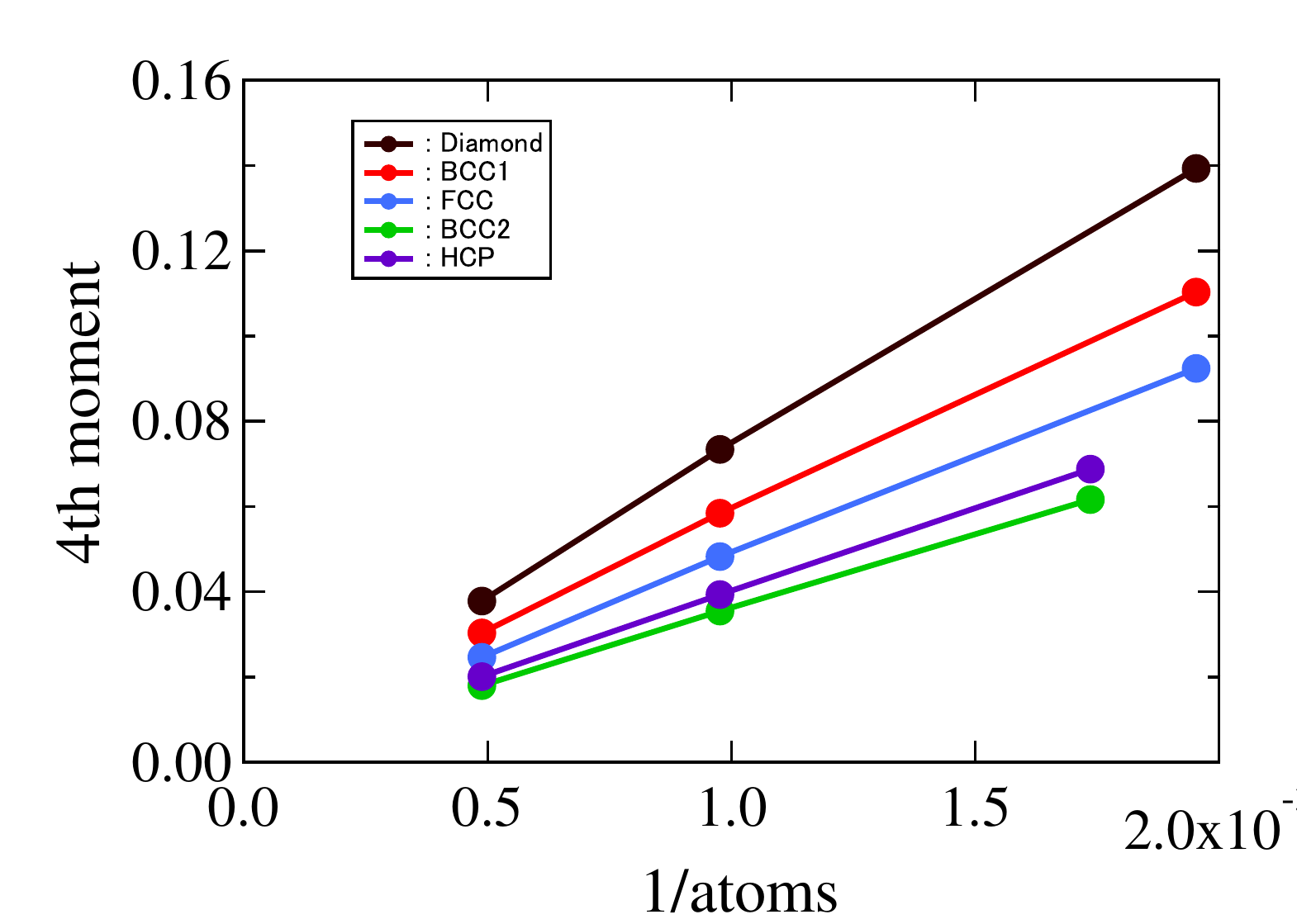}
        \end{center}
      \end{minipage}
    \end{tabular}
    \caption{2nd, 3rd, and 4th order moments of DOEs for CFS along the inverse of the number of atoms. These certainly indicate that moments has inverse proportion to the number of atoms.}
    \label{fig.rev_moment_all}
  \end{center}
\end{figure*}

For confirming the universality of density of states in CFS, we construct the matrix from the composition-fixed system (i.e., an equiatomic system) with MC simulation.
In previous study for FCC lattice with random matrix \cite{EMRS5}, we have showed that statistical dependence of density of microscopic states is gradually eliminated when the number of atoms, $N$, increases.
Therefore, we demonstrate whether this tendency of $N$ dependence can be holed in set1 and set2. 
Figures.~\ref{fig.doe_all} shows the landscape of DOEs for set1 and set2.
In terms of decrease of sub-peaks and location of highest-peak, we can easily understand that statistical interdependence is confirmed with the increase of $N$, which meets our previous research \cite{EMRS5}.
For more quantitative comparison, we estimate 2nd-4th moments of DOEs defined above, and show Figs.~\ref{fig.moment_all} and \ref{fig.rev_moment_all};
Figure~\ref{fig.moment_all} clearly shows that when the increase of the size of spatial constraint, all the moments for practical system become close to that for RM, and Fig.~\ref{fig.rev_moment_all} shows that all moments for practical system may be in inverse proportion to $N$;we derive the model showing this connection of 2nd moment followings, and that of 3rd and 4th moment in Appendix B.\\
\ First, we note that $R$ defined in Eq.~(\ref{eq:matrix}) is symmetric matrix, which leads $(s,t)$ element of $R^2$ to be equal to $(r_{st})^2$.
From the linear algebra, 2nd moment defined in this paper, $M_2$, can be represented as
\begin{eqnarray}
M_2&=&\sqrt{\frac{1}{m}\left(\sum_{(i,j)}(r_{ij})^2+\sum_i(r_{ii})^2\right)-M_{1}^2} .,
\label{eq_2}
\end{eqnarray} 
where ${(i,j,k,\cdot\cdot\cdot)}$ denotes the combination whose all elements are different. 
We normalize each column of $A$ with its variance taking 1, which develop Eq.~(\ref{eq_2}) as
\begin{eqnarray}
M_2&=&\sqrt{\frac{1}{m}\sum_{(i,j)}(r_{ij})^2} .
\label{eq_3}
\end{eqnarray} 
In order to proof the inverse proportion, Eq.~(\ref{eq_2}) tells that we need to show the $r_{ij}$ has the inverse proportion to $N$.
We note that $r_{ij}$ is covariance of $q_i/{\langle}q_i\rangle_{sd}$ and $q_j/{\langle}q_j\rangle_{sd}$ (hereinafter, defined as $\acute{q}_i$ and $\acute{q}_j$).
With average of each column of $A$ taking 0, we obtain 
\begin{eqnarray}
r_{ij}&=&\int\int g(\acute{q}_i, \acute{q}_j)(\acute{q}_i-{\langle}\acute{q}_i\rangle_1)(\acute{q}_j-{\langle}\acute{q}_j\rangle_1)d\acute{q}_i d\acute{q}_j \nonumber\\ 
&=&\int\int g(\acute{q}_i,\acute{q}_j )\acute{q}_i \acute{q}_jd\acute{q}_i d\acute{q}_j.
\label{eq_4}
\end{eqnarray}
Here, we consider how many spin products consisting of $q_i$, $\displaystyle \prod_{i{\in}u} \sigma_i$, is changed with the change of $\acute{q}_i$, $\delta\acute{q}_i$:
\begin{equation}
 {\delta}\acute{q}_i=\frac{{\delta}C_i/Ns}{{\langle}q_i\rangle_{sd}}=\frac{{\delta}C_i}{Ns\cdot{\langle}q_i\rangle_{sd}},
\label{eq_5}
\end{equation}
where ${\delta}C_i$ denotes the the number of spin products changed, $s$ denotes the number of spin products per site.
Eq.~\ref{eq_5} shows that under the same ${\delta}C_i$, $\delta\acute{q}_i$ is in inverse proportion to $N$ because ${\langle}q_i\rangle_{sd}=1/\sqrt{Nd}$ \cite{hensa}\cite{CDOS}. 
With using this connection, when we consider how many spin products consisting of $q_i$ is obligatorily changed by the change of $q_j$, which we define as $H\left( {\delta}\acute{q}_i, {\delta}\acute{q}_j\right)$, $H\left( {\delta}\acute{q}_i, {\delta}\acute{q}_j\right)$ should be 
\begin{equation}
H\left({\delta}q_i,{\delta}q_j\right)=\frac{1}{N}h\left({\delta}q_i,{\delta}q_j\right),
\end{equation}
where $h\left({\delta}q_i,{\delta}q_j\right)$ denotes ideal $H\left({\delta}q_i,{\delta}q_j\right)$ at $N=1$, from the view point of the 2D space of  spin products for $q_i$ and $q_j$.
With $H$, we introduce the model to represent $g({\delta}q_i,{\delta}q_j)$ as
\begin{eqnarray}
g({\delta}q_i,{\delta}q_j)&\simeq&g_i({\delta}q_i)g_j({\delta}q_j)\left(1+H\left({\delta}q_i,{\delta}q_j\right)+{\Delta}\left({\delta}q_i,{\delta}q_j\right)\right). \nonumber \\
\label{eq_7}
\end{eqnarray} 
With this model, ${\int}g({\delta}\acute{q}_i) {\delta}\acute{q}_id{\delta}\acute{q}_i=0$ because the landscape of DOEs certaily indicate  $g({\delta}\acute{q}_i)$ should be Gaussian distribution:
Therefore, $r_{ij}$ should be inverse proportion to $N$:
We can demonstrate the inverse proportion of moments to N.
This connection certainly indicate that statistical dependence can be vanished with taking limit of $N$ even in any lattices.
Consequently, we can confirm the universality of density of states in CFS, which strongly support our PS approach.

\subsection{Universality of density of states in CUFS}
\begin{figure*}[p]
  \begin{center}
    \begin{tabular}{c}
%
      \begin{minipage}{0.33\hsize}
        \begin{center}
          \includegraphics[width=\linewidth]{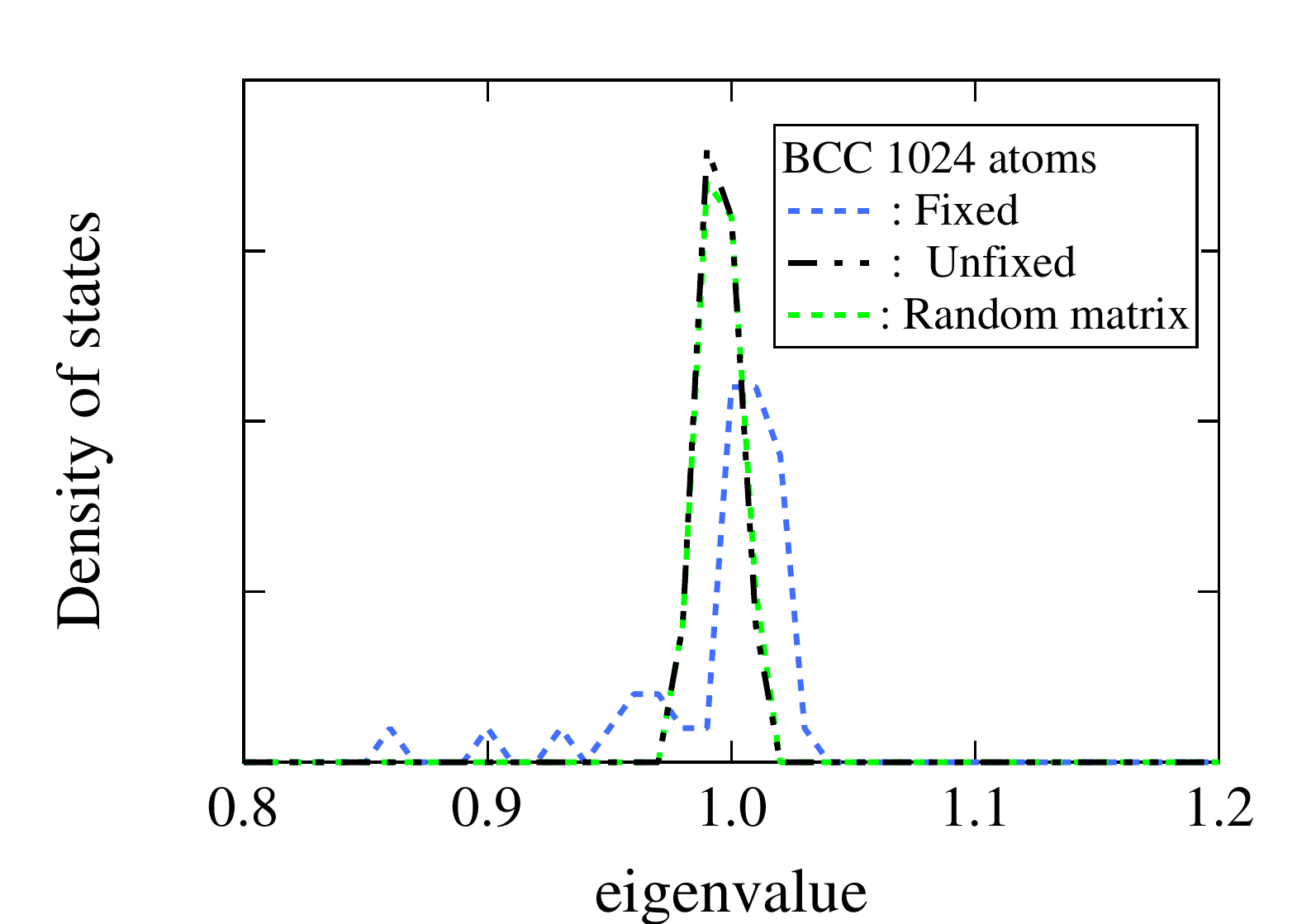}
        \end{center}
      \end{minipage}
      \begin{minipage}{0.33\hsize}
        \begin{center}
          \includegraphics[width=\linewidth]{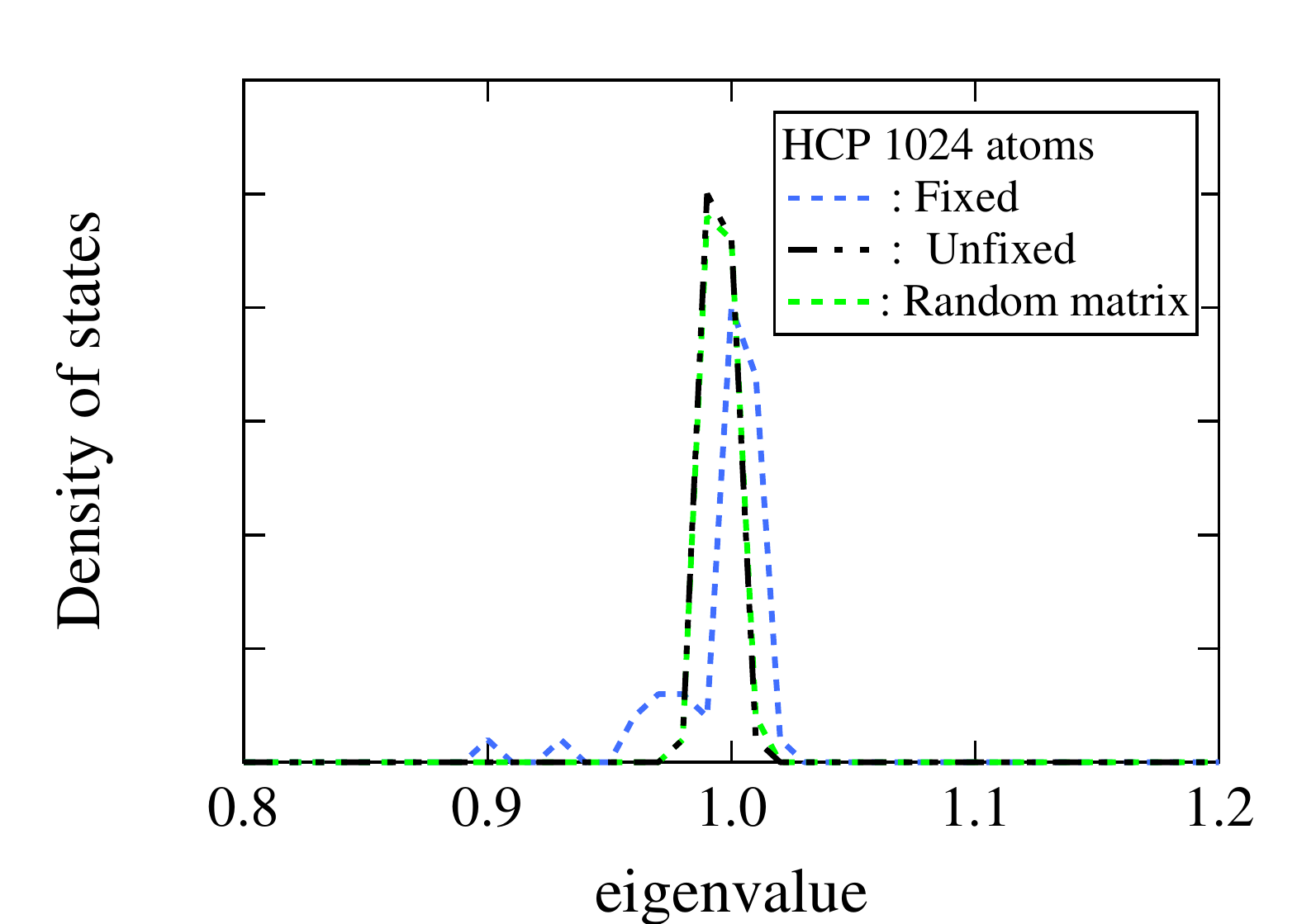}
        \end{center}
      \end{minipage}
	\begin{minipage}{0.33\hsize}
        \begin{center}
          \includegraphics[width=\linewidth]{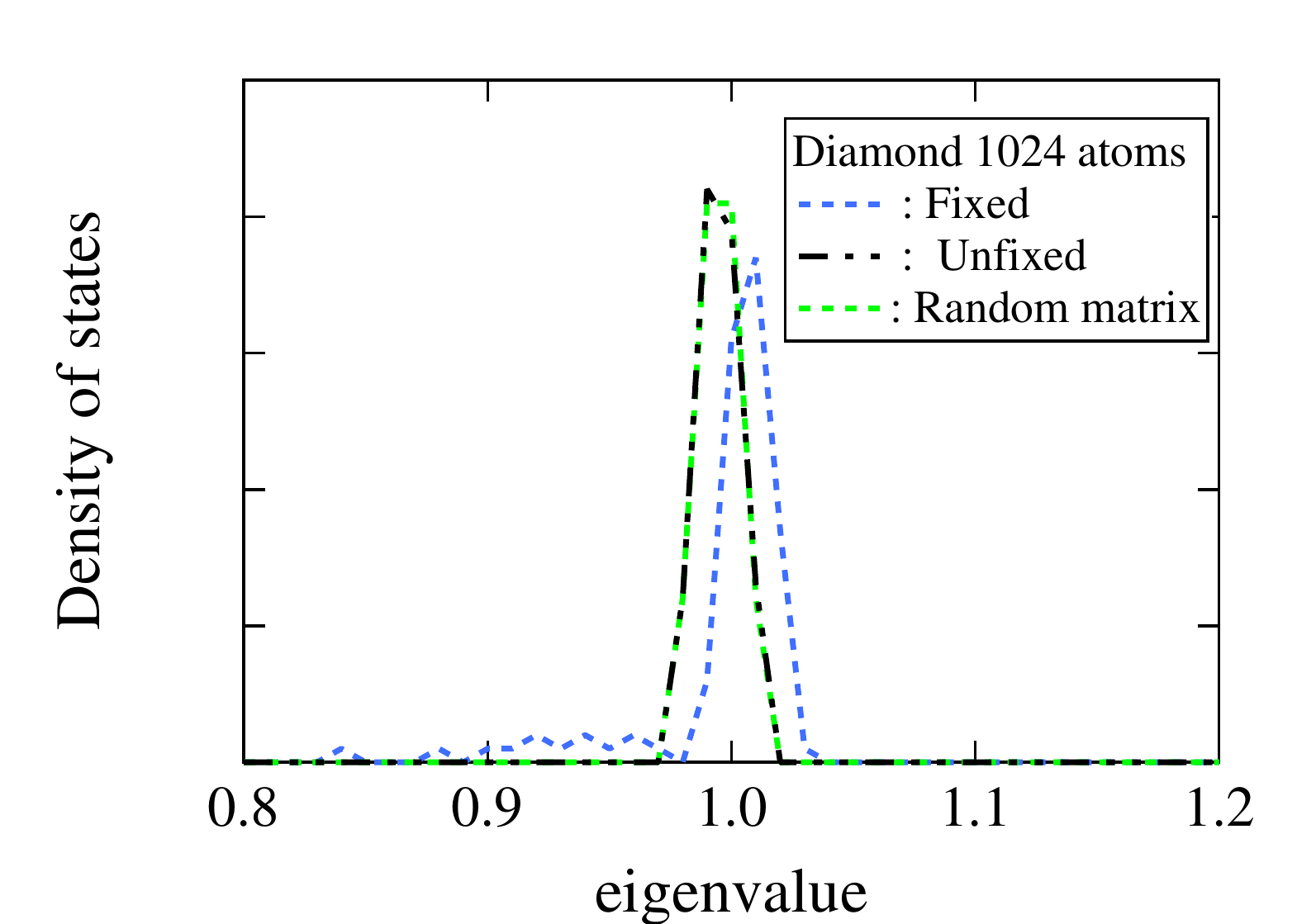}
        \end{center}
      \end{minipage}
 \end{tabular}
    \caption{ Density of states along eigenvalues of covariance matrix (DOE), constructed from  CFS, CUFS, and RM. These shows that DOE for CUFS is more similar to that for RM than for CFS in every lattice.}
    \label{fig.doe_fix_unfix_all}
\end{center}
\end{figure*}

\begin{figure*}
  \begin{center}
    \begin{tabular}{c}
%
      \begin{minipage}{0.33\hsize}
        \begin{center}
          \includegraphics[width=\linewidth]{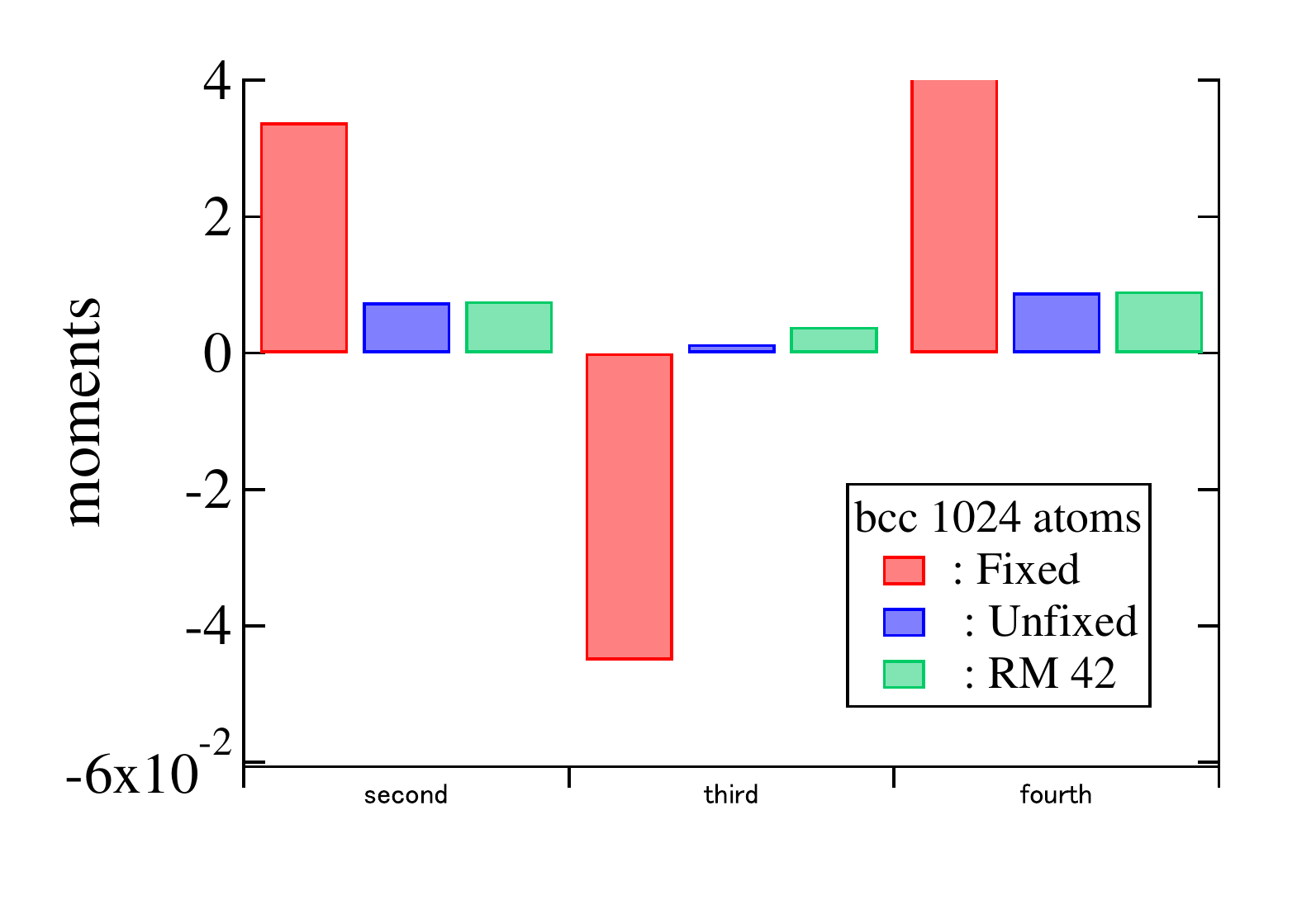}
        \end{center}
      \end{minipage}
      \begin{minipage}{0.33\hsize}
        \begin{center}
          \includegraphics[width=\linewidth]{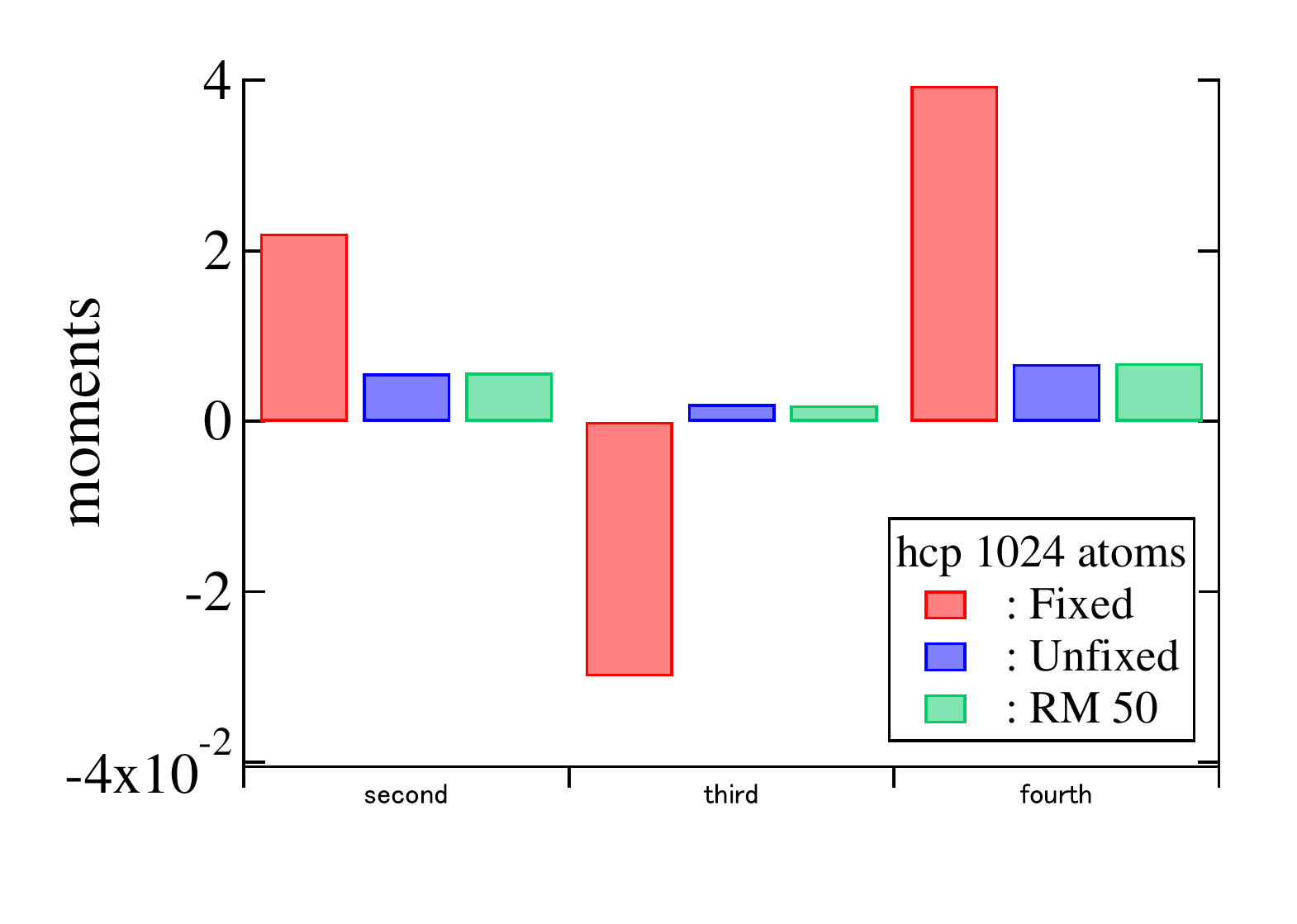}
        \end{center}
      \end{minipage}
	\begin{minipage}{0.33\hsize}
        \begin{center}
          \includegraphics[width=\linewidth]{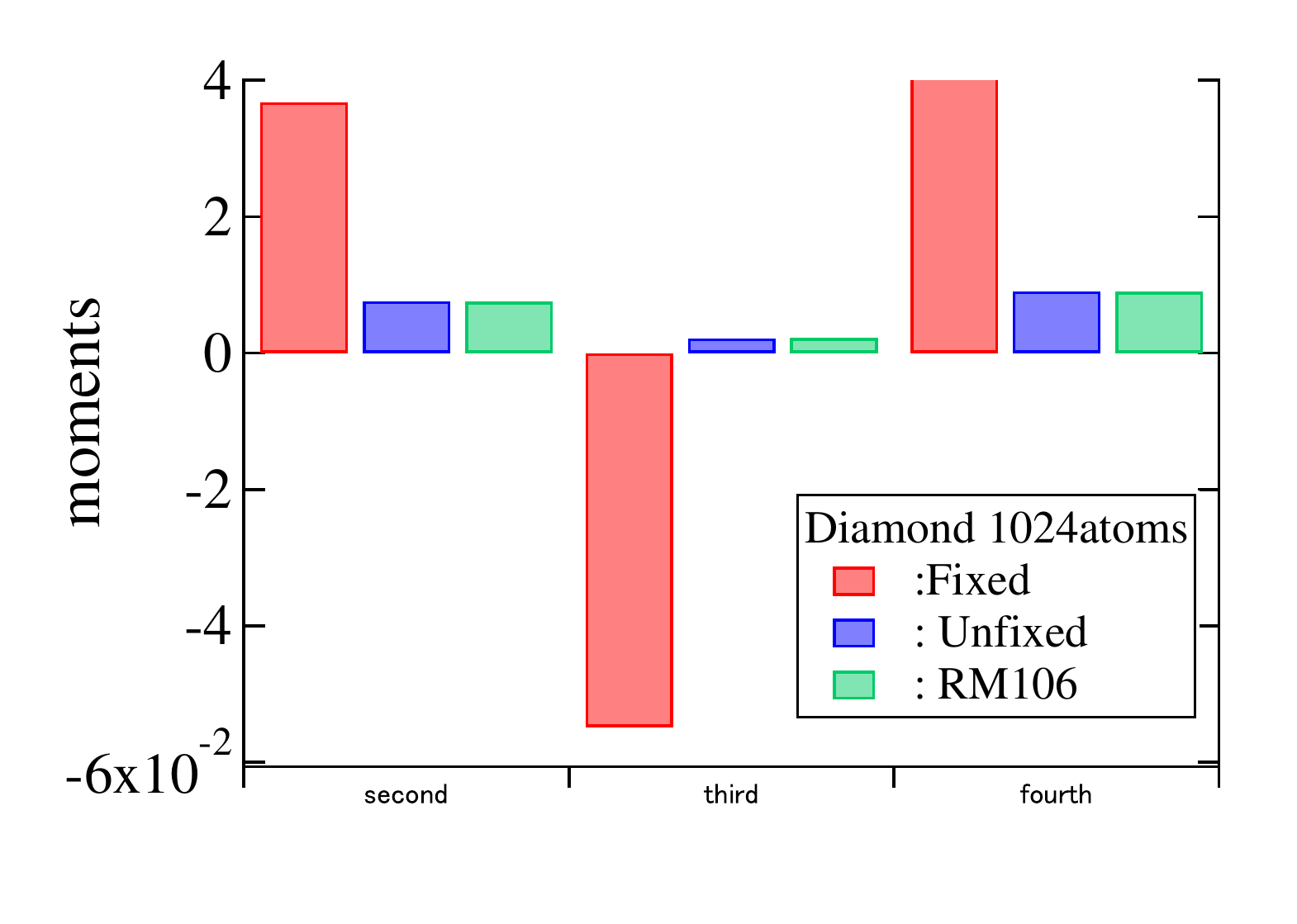}
        \end{center}
      \end{minipage}
 \end{tabular}
    \caption{2nd, 3rd, and 4th-order moments of DOEs for CFS, and CUFS, and RM.
        These show excellent agreement of RM and CUFS quantitatively.  .}
    \label{fig.moment_fix_unfix_all}
\end{center}
\end{figure*}

In CUFS, we have shown that statistical interdependence is more eliminated than in CFS under the same $N$ in FCC lattice; even in about 1000 atoms, the moments of CUFS successfully agree with that of RM \cite{EMRS4}.
In this section, therefore, we confirm whether this tendency of CUFS remains in any lattice in order to demonstrate vanishment of statistical interdependence in CUFS.
\\
\ In order to construct the matrix from CUFS with MC simulation, we have introduced a local system in an ideally large composition-fixed system; we can theoretically determine the probability of composition of local system from the composition of an ideally large system with binomial distribution.
Therefore, based on this probability, we set the sampling times for the each composition of local system in set1 and set2.
\\
Figs.~\ref{fig.doe_fix_unfix_all}  shows the landscape of DOEs, which shows statistical independence is more confirmed in CUFS under the same $N$.
Moreover, Fig.~\ref{fig.moment_fix_unfix_all} shows the moments of DOEs for each lattices, which clarify the moments of CUFS shows excellent agreement with that of RM in every simulation.
These clearly meets our previous research, and the tendency in FCC lattice remains.
Moreover these two results of set1 and set2 indicate however we make SL from ML, statistical independence is confirmed, which means statistical independence in any lattices
Consequently, we can confirm the universality of density of states in CUFS, which strongly support our GPS approach.   

\section{Conclusion}
 In this study, we confirm the universality of density of microscopic states in non-interacting system; this means statistical interdependence is vanished in any lattices even though the basis functions themselves are not essentially statistically independent.
This enable one to obtain information of configuration of solute atoms, free energy, phase diagram with performing first-principles calculation on few special microscopic states combined with our established theory.
Moreover, this study can open the door to new approach based on statistical independence.

\section*{Acknowledgement}
This work was supported by a Grant-in-Aid for Scientific Research (16K06704), and a Grant-in-Aid for Scientific Research on Innovative Areas “Materials Science on Synchronized LPSO Structure” (26109710) from the MEXT of Japan, Research Grant from Hitachi Metals$\cdot$Materials Science Foundation, and Advanced Low Carbon Technology Research and
 Development Program of the Japan Science and Technology Agency (JST).

\setcounter{equation}{0}
\setcounter{figure}{0}
\setcounter{table}{0}
\setcounter{section}{0}
\appendix
\renewcommand{\theequation}{A\arabic{equation}}
\renewcommand{\thefigure}{A\arabic{figure}}
\renewcommand{\thetable}{A\arabic{table}}
\section{Table}
 In this section, we show the condition of experiment, such as the detail of how we make SL from ML and what kind of clusters we consider in each set, and the difference between BCC1 and BCC2.
Figure.~\ref{HSL_LSL} shows how we make SL from ML.
In ML of set 1 (FCC lattice), clusters considered are pair clusters up to 6NN, triplet clusters consisting of up to 6NN pairs, resulting in 29 basis functions(i.e., $n$=29). We sample 500,000 microscopic states (i.e., $m=500,000$) by performing MC simulation for the MC-cells.
The condition of the study of SL, BCC1 and Diamond, is decided based on this FCC setting (explained above in Sec.~\ref{sec:level2}); 6NN pair in FCC is extended to 8NN pair in BCC1 and 12NN in Diamond.
 Therefore, we do calculation with Table.~\ref{condition_set1}

\begin{figure*}
  \begin{center}
    \begin{tabular}{c}
%
      \begin{minipage}{0.33\hsize}
        \begin{center}
          \includegraphics[width=\linewidth]{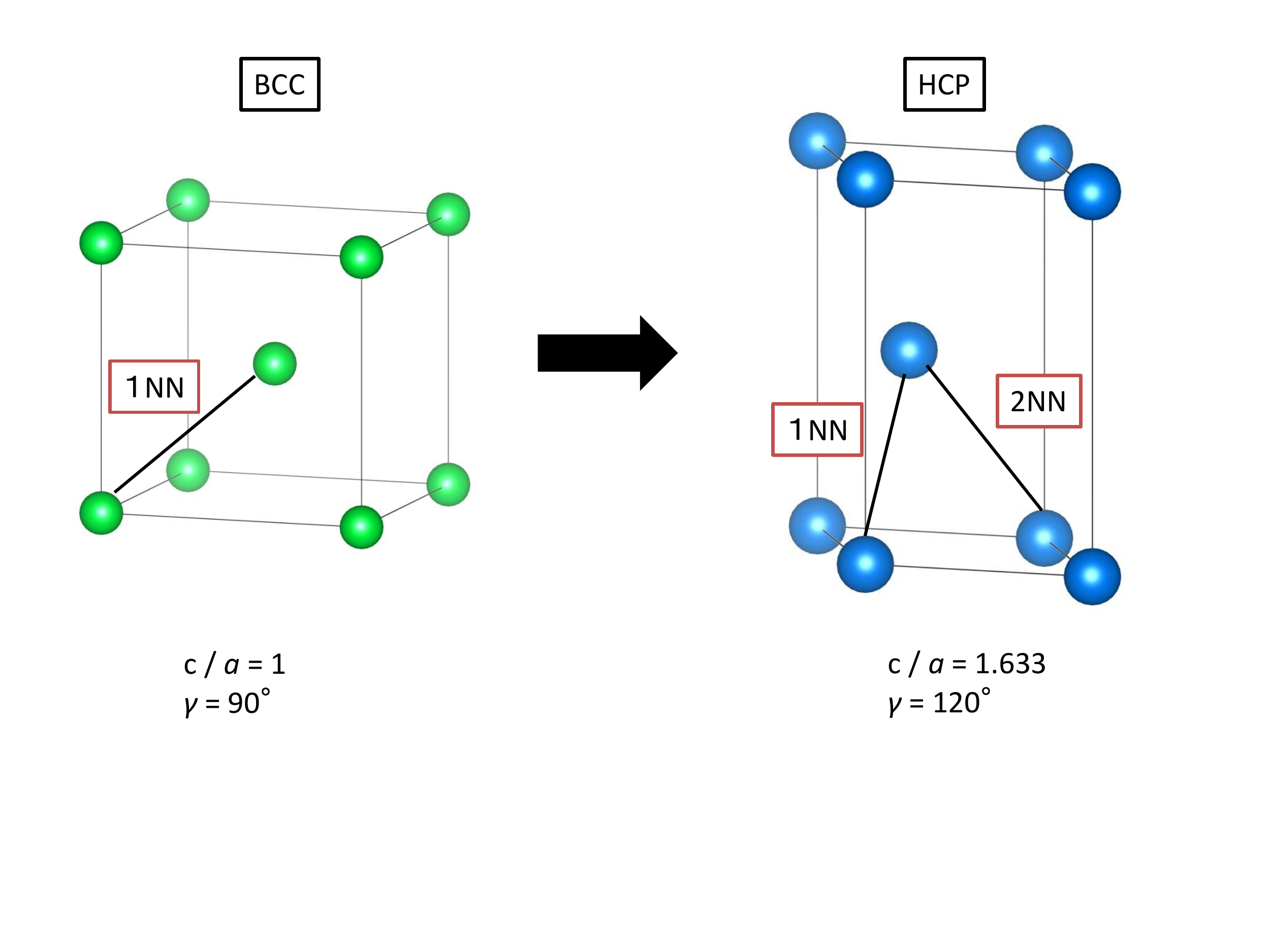}
        \end{center}
      \end{minipage}
      \begin{minipage}{0.33\hsize}
        \begin{center}
          \includegraphics[width=\linewidth]{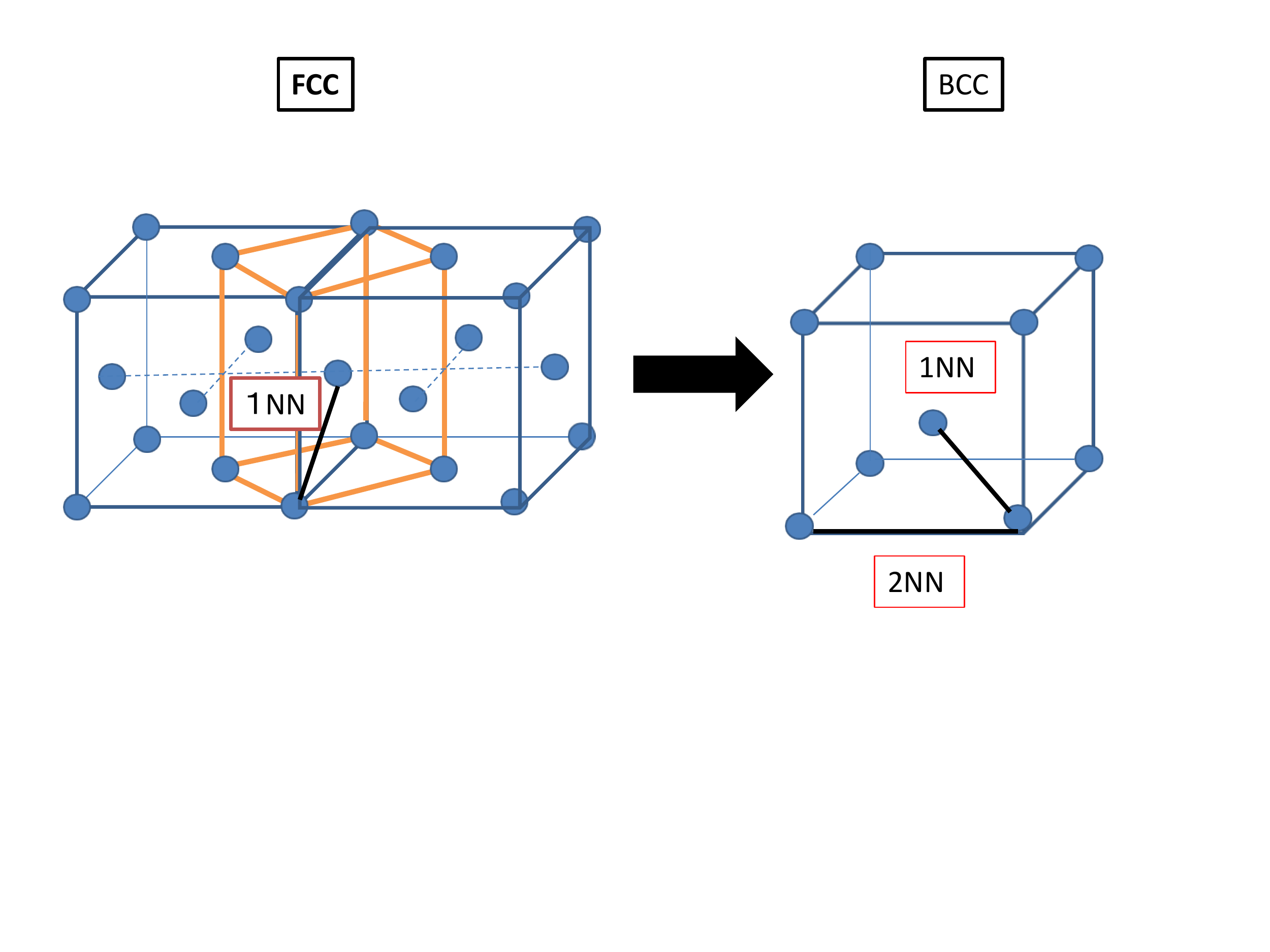}
        \end{center}
      \end{minipage}
	\begin{minipage}{0.33\hsize}
        \begin{center}
          \includegraphics[width=\linewidth]{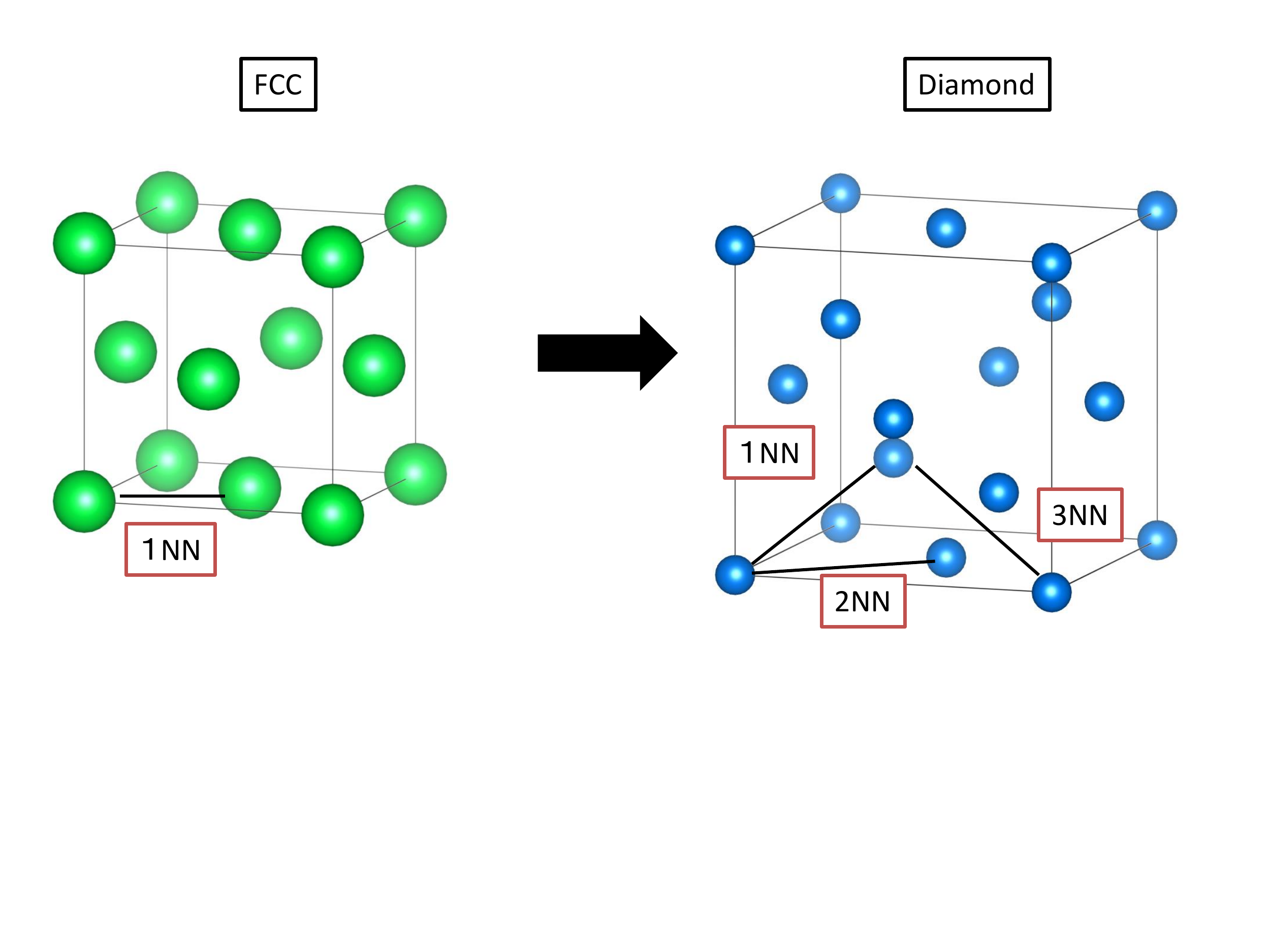}
        \end{center}
      \end{minipage}
 \end{tabular}
    \caption{ These figures shows how we construct SL from ML.}
    \label{fig.HSL_LSL}
\end{center}
\end{figure*}
\begin{table}[h]
  \begin{center}
         \begin{tabular}{|l|l|l|l|}\hline 
         &FCC & BCC1& Diamond \\ \hline
         Basis functions considered&29&42&106 \\ \hline
         Sampling times&500,000&724,138&1,827,586\\ \hline
         Number of atoms &\multicolumn{3}{c|}{512, 1024, 2048atoms}\\ \hline 
         
            \end{tabular}
         \caption{Condition of set1.}
         \label{condition_set1}
  \end{center}
\end{table}
\begin{table}[h]
  \begin{center}
         \begin{tabular}{|l|l|l|}\hline 
         &BCC2 & HCP \\ \hline
         Clusters considered&17&50 \\ \hline
         Sampling times&500,000&1470588\\ \hline
         Number of atoms &\multicolumn{2}{c|}{576, 1024, 2048 atoms}\\ \hline 
            \end{tabular}
         \caption{Condition of set2.}
         \label{condition_set2}
  \end{center}
\end{table}
Also in set2, first we determine the condition of ML, BCC2; pair clusters considered up to 5NN, triplet clusters consisting of up to 5NN pairs, resulting in 17 basis functions(i.e., $n$=17). We sample 500,000 microscopic states (i.e., $m=500,000$) with MC simulation, which decide condition of SL, HCP.
In HCP, we consider pair clusters up to 10NN (except 6NN and 9NN), and triplet clusters up to 10NN (also except 6NN and 9NN) pairs.
Figure.~\ref{condition_set2} shows the condition of set2 more in detail.
\renewcommand{\theequation}{B\arabic{equation}}
\section{Connection of 3rd/4th moment and the number of atoms}
 In this section, we show the proof of inverse proportion of 3rd/4th moment  to $N$, which is seen in Fig,~\ref{fig.rev_moment_all}.
 The difference from 2nd moments is Eqs.~(\ref{eq_2}) and~(\ref{eq_2}): the inverse proportion ,to $N$, of $r_{ij}$  at $i{\neq}j$ of course remains when we consider 3rd/4th moment.
Therefore, we can represent 3rd/4th moment with simple representation as
\begin{eqnarray}
M_3&=&\sqrt[3]{\sum_{t=0}^3{a_{3t}N^{-t}}}.
\label{eq_B1}\\
M_4&=&\sqrt[4]{\sum_{t=0}^4{a_{4t}N^{-t}}}.
\label{eq_B2}
\end{eqnarray}
With this representation, we can easily obtain the inverse proportion of 3rd/4th moment to $N$ at large $N$.

\end{document}